\documentclass[11pt,twoside]{article}

\usepackage{asp2006}
\usepackage{fancyhdr,graphicx,caption,rotating,multirow,lscape}
\DeclareGraphicsExtensions{.eps,.eps.gz,.ps,.jpg,.tiff}

\begin{document}

\title{Exoplanets and the Rossiter-McLaughlin Effect}
\author{J.~N.~Winn}
\affil{Massachusetts Institute of Technology, Cambridge, MA, USA}

\begin{abstract} 

  A transiting planet eclipses part of the rotating stellar surface,
  thereby producing an anomalous Doppler shift of the stellar
  spectrum.  Here I review how this ``Rossiter-McLaughlin Effect'' can
  be used to characterize exoplanetary systems. In particular, one can
  measure the angle on the sky between the orbital axis and the
  stellar rotation axis. This may help to discriminate among migration
  theories. Measurements have been made for 4 exoplanets, and in all
  cases the spin and orbital axes are fairly well-aligned.  In the
  future, the Rossiter-McLaughlin effect may also be important as an
  alternative means of probing exoplanetary atmospheres, and for
  confirming the transits of objects identified by the satellite
  missions {\it Corot} and {\it Kepler}.

\end{abstract}

\section{Introduction}

For most of the participants in this workshop, the word ``transit''
brings to mind an image such as the left panel of Fig.~1.  This is the
beloved transit light curve, an inverted boxcar with its corners
sanded down by limb darkening.  For those of us who study the
Rossiter-McLaughlin effect, the object of our affection is shown in
the right panel of Fig.~1: an elegant antisymmetric blip with a gently
sloping baseline.

\begin{figure}[!ht]
\centering
\includegraphics[angle=0,width=13.5cm]{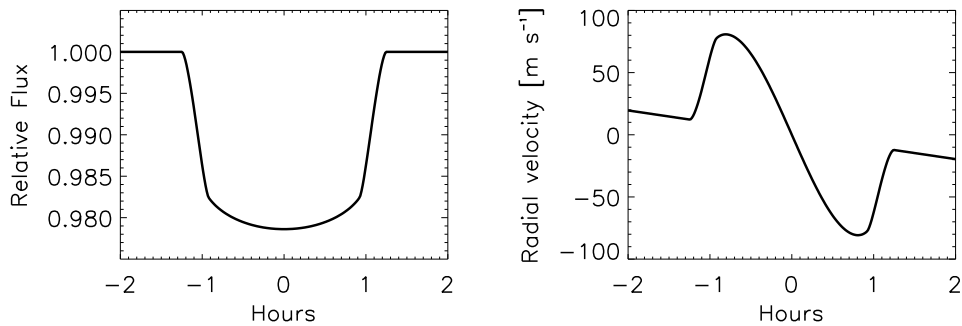}

\caption{Simulation of a photometric transit (left) and the
corresponding spectroscopic transit (right).  The system parameters
were chosen to be similar to those of TrES-1 (Alonso et al.~2004; Winn,
Holman, \& Roussanova 2006).}
\end{figure}

\clearpage

What is the meaning of this wiggly waveform?  It is the variation in
the apparent Doppler shift of the star throughout a transit, the most
prominent {\it spectroscopic} effect of the planet's passage. It
arises because of stellar rotation. The emergent spectrum from a given
point on the stellar disk is Doppler-shifted by an amount that depends
on the local line-of-sight velocity. The spread in velocities across
the disk broadens the spectral lines (along with thermal and turbulent
broadening). When the planet hides a portion of the stellar surface,
the corresponding velocity components are missing from the spectral
lines. This distortion is usually manifested as an ``anomalous''
Doppler shift. When the planet is in front of the approaching
(blueshifted) half of the stellar disk, the starlight appears slightly
redshifted. The anomalous Doppler shift vanishes when the planet is in
front of the stellar rotation axis, and then reverses sign as the
planet moves to the receding (redshifted) half of the stellar disk.

This phenomenon is called the ``Rossiter-McLaughlin effect,'' in honor
of the two gentlemen who described it in a back-to-back pair of papers
in the Astrophysical Journal (Rossiter 1924; McLaughlin 1924),
although in fact the effect had been observed years earlier (Forbes
1911; Schlesinger 1911). Of course, those observations involved
eclipsing binary stars, rather than exoplanets.

The exoplanetary Rossiter-McLaughlin (RM) effect was first observed by
Queloz et al.~(2000) and Bundy \& Marcy~(2000) during transits of
HD~209458b. It has since been observed in at least 3 other systems,
and the motivation for additional measurements is strong. In this
contribution, I explain the motivation, review the existing
measurements and their implications, and discuss prospects for future
observations.  More details on the theory can be found in the works by
Ohta, Taruya, \& Suto (2005), Gimenez (2006), and Gaudi \& Winn
(2007).

\section{Spin-Orbit Alignment}

By observing the RM effect with a high signal-to-noise ratio, one can
determine the trajectory of the planet relative to the (sky-projected)
stellar rotation axis. Specifically, one can measure the angle
$\lambda$ between the sky projections of the orbital axis and the
stellar rotation axis. This is illustrated in Fig.~2. Pictured are
three trajectories of a transiting planet, all of which have the same
impact parameter (and hence produce exactly the same photometric
signal), but which differ in $\lambda$ (and hence produce different RM
waveforms).

Since the angular momentum of the star and of the orbits are derived
from the same source---the protostellar disk---one would naturally
expect the spin and orbit to be well-aligned and $\lambda$ to be
small.  Indeed, in the Solar system, the planetary orbital axes are
aligned with the Solar rotation axis within $\sim$10$^{\circ}$. Why,
then, would one bother measuring $\lambda$ for exoplanets?


\begin{figure}[!ht]
\centering
\includegraphics[angle=0,width=13.5cm]{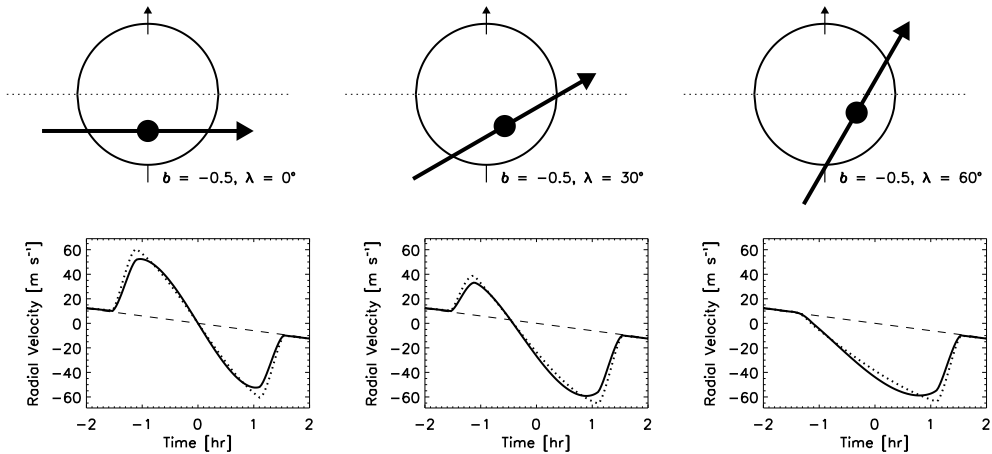}
\caption{The dependence of the RM waveform on $\lambda$, from Gaudi \&
Winn (2007). Three different trajectories of a transiting planet are
shown along with the corresponding RM waveform. Solid lines include
the effect of limb darkening; dotted lines neglect limb darkening.}
\end{figure}

An answer that some readers may find satisfactory is ``what one can
measure, one should measure.'' Exoplanets have rewarded observers with
surprises in the past. Those readers requiring a theory-based
motivation might ask whether or not the migration mechanism for hot
Jupiters preserves spin-orbit alignment. Migration via tidal
interactions with a disk would not be expected to perturb spin-orbit
alignment, and may even drive the system toward closer alignment (see,
e.g., Ward \& Hahn 1994, 2003). In contrast, migration mechanisms
involving disruptive events such as planet-planet interactions or
planetesimal collisions would act to enhance any initial
misalignment. Another migration theory involves the Kozai mechanism,
in which a companion star causes oscillations in the planetary orbit's
eccentricity and inclination. By the time tides circularize the orbit
and halt ``Kozai migration,'' the orbital inclination can change
substantially (Wu \& Murray 2003; Eggenberger, Udry, \& Mayor 2004;
D.\ Fabrycky \& S.\ Tremaine, priv.\ comm.). Thus, measuring
spin-orbit alignment offers a possible means for discriminating among
migration theories, or at least for identifying particular planets
that migrated through disruptive mechanisms.

Results for $\lambda$ have been published for two systems. For
HD~209458, the latest result is $\lambda=-4.4^{\circ} \pm 1.4^{\circ}$
(Winn et al.~2005). The small but nonzero angle is reminiscent of
Solar system planets. For HD~189733, the result is also a very close
alignment: $\lambda = -1.4^{\circ} \pm 1.1^{\circ}$ (see Fig.~3, from
Winn et al.~2006). In addition, a paper in press by Wolf et al.~(2007)
states $\lambda = 11^{\circ} \pm 15^{\circ}$ for HD~149026. The lower
accuracy in that case is mainly due to the smaller size of the planet
relative to the star. Most recently, the TrES-1 system was found to be
consistent with $\lambda=0$ within about $30^{\circ}$ (N.~Narita, this
volume).


Together, these results rule out the (admittedly rather extreme)
hypothesis of completely random alignment, with $>$99.9\% confidence.
Apparently, in these systems at least, the migration mechanism
preserved spin-orbit alignment. (The observed alignment probably
reflects the initial condition, because the timescale for tidal
coplanarization is very long; see Winn et al.~2005.)  Further
measurements are needed to estimate the actual distribution of
$\lambda$, and of course the discovery of even a single example of a
grossly misaligned system would be of great interest. For planning
purposes, Gaudi \& Winn (2007) have provided formulas that can be used
to estimate the accuracy with which $\lambda$ can be measured, given
the geometry of the system and the characteristics of the data.

It is important to remember that $\lambda$ is the angle between the
{\it projected} orbital and rotation axes. The inclinations of those
two axes with respect to the sky plane must be determined using other
means. The orbital inclination can be determined from the transit
light curve, but in general the stellar inclination is unknown. For
the special case of HD~189733, however, the rotation period has been
measured; the star is chromospherically active and exhibits
quasiperiodic flux variations, presumably due to star spots. The
combination of a measured rotation period, stellar radius, and $v\sin
i$ places a constraint on $i$. Hence this is the first case for which
the true (3-d) angle between the orbital and rotational axes can be
measured, and the result is an upper bound of $27^{\circ}$ with 95\%
confidence (Winn et al.~2007).

\begin{figure}[!ht]
\centering
\includegraphics[angle=0,width=13.5cm]{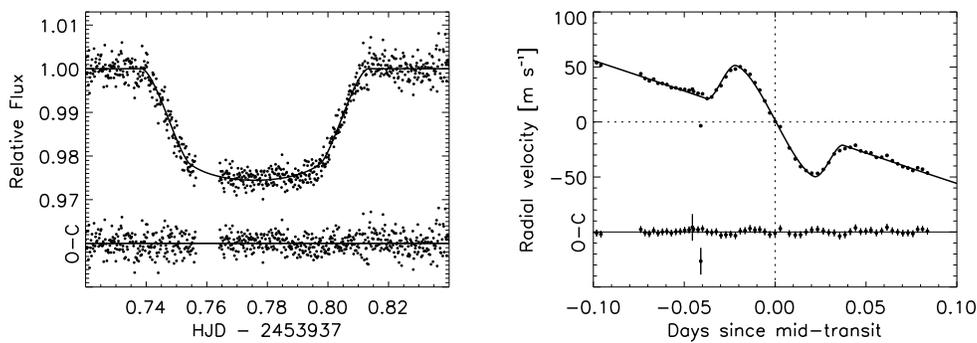}
\caption{The photometric and spectroscopic transit of
HD~189733, from Winn et al.~(2006). Photometry was obtained
with the 1.2~m telescope at the Fred L.~Whipple Observatory,
using Keplercam and a Sloan $z$ filter. Radial velocities
were derived from spectra that were measured
with the Keck~I~10~m telescope and HIRES. The residuals
(O$-$C) are displayed beneath each data set.}
\end{figure}

\section{Transmission Spectroscopy}

Snellen (2004) realized that the RM effect offers an alternative
method of ``transmission spectroscopy.'' During a transit, a small
portion of the received starlight is filtered through the planetary
atmosphere, which may imprint detectable absorption (or emission)
features. At the wavelength of a strong absorption line, the effective
radius of the planet is larger. This can be detected through the
wavelength-dependence of the photometric transit depth, a technique
that was used by Charbonneau et al.~(2002) to detect atomic sodium in
the atmosphere of HD~209458. Since the RM anomaly also depends on the
effective radius of the planet, the wavelength-dependence of the RM
effect can be used for transmission spectroscopy. Rather than relying
on accurate time-series photometry, the RM method relies on the
comparison of the Doppler shifts of different lines within a single
spectrum. In principle, this could lead to more accurate results (at
least in comparison to ground-based photometry), although only upper
limits have been achieved to date (Snellen~2004).

\section{Transit Confirmation}

It is interesting to compare the velocity amplitude $K_O$ of the
star's orbital motion with the velocity amplitude $K_R$ of the RM
effect. For a small planet of mass $M$ on an edge-on circular orbit
with period $P$, Gaudi \& Winn (2007) showed that the order of
magnitude of $K_R/K_O$ is
\begin{displaymath}
\frac{K_R}{K_O}
\sim 0.3 \left(\frac{M}{M_{\rm Jup}}\right)^{-1/3}
\left(\frac{P}{3~{\rm days}}\right)^{1/3}
\left(\frac{v \sin i}{5~{\rm km~s}^{-1}}\right).
\end{displaymath}
Thus, for hot Jupiters, the anomalous velocity is smaller than the
orbital velocity. However, for smaller planets with longer periods,
the amplitude of the RM effect will {\it exceed} the stellar orbital
velocity. For an Earth-mass planet with a period of one year, $K_R/K_O
\sim 3$ for $v\sin i=5$~km~s$^{-1}$.

This raises the appealing possibility of using the RM effect to
confirm transits that will be detected by the forthcoming satellite
missions {\it Corot} and {\it Kepler}. The most exciting discoveries
by these satellites will be very small planets with transit depths of
$\sim$$10^{-4}$ or less. One would like to detect the spectroscopic
orbit and thereby learn the planetary mass, but this will be
challenging because $K_O$ is only $\sim$10~cm~s$^{-1}$ for an
Earth-like planet in the habitable zone of a solar-type star. It would
be useful to have a means of confirming the transits before chasing
after the spectroscopic orbit. Photometric confirmation from the
ground may prove very difficult. RM confirmation appears more feasible
for at least some stars, not only because $K_R > K_O$ as mentioned
above, but also because the RM velocity variation occurs over the time
scale of the transit duration ($\sim$1~day), which is much shorter
than the time scale of the orbital velocity variation
($\sim$1~yr). Both of these points are illustrated in Fig.~4.

\begin{figure}[!ht]
\centering
\includegraphics[angle=0,width=13.5cm]{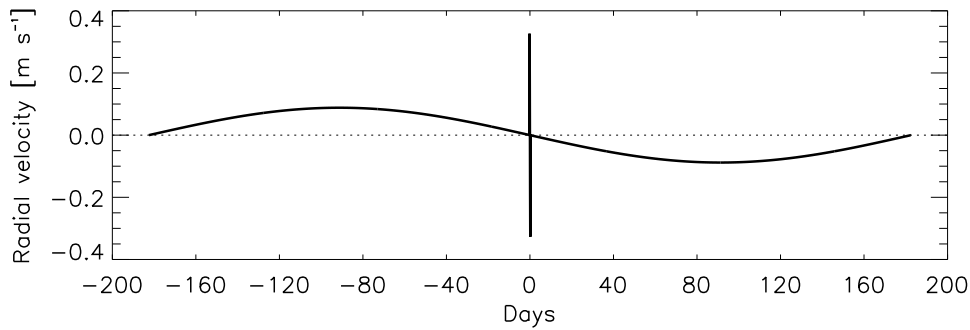}
\caption{Simulated spectroscopic signal of a transiting terrestrial
planet in the habitable zone of a solar-type star with $v\sin i =
5$~km~s$^{-1}$, from Gaudi \& Winn (2007). A circular, edge-on orbit
is assumed. The sinusoid with a period of 1~yr is the spectroscopic
orbit.  The spike near time zero is the RM effect, which occurs over
$\sim$1~day.}
\end{figure}

\clearpage

This idea is discussed at greater length in these proceedings by
W.~Welsh.  A related idea by Ohta, Taruya, \& Suto~(2006) is to search
for planetary rings using RM observations. Here too, the RM effect is
used as an alternative means of measuring the transit depth, relying
on the measurement of spectral features rather than photometric
stability.

\section{Summary}

The RM effect is the anomalous Doppler shift of starlight that is
observed during a planetary transit, due to stellar rotation. It
provides another fundamental observable for exoplanetary systems: the
degree of alignment between the planetary orbital axis and the stellar
rotation axis (in projection on the sky). Observations of the RM
effect are likely to grow in importance in the future, as an
alternative means of confirming photometric transits and possibly also
for transmission spectroscopy.

\vspace{0.5cm}

\acknowledgements

I have enjoyed the privilege of working with Scott Gaudi on RM theory
and John Johnson on RM observations; the results described here owe
much to them. I am grateful to Yasushi Suto for stimulating my
interest in this subject and for hospitality in Tokyo. Dave
Charbonneau, Geoff Marcy, Bob Noyes, Dimitar Sasselov, and Ed Turner
have also provided helpful advice. I thank Cristina Afonso and the
other organizers for staging a wonderful meeting in Heidelberg.



\begin{thebibliography}{}

\bibitem[Alonso et al.(2004)]{2004ApJ...613L.153A} Alonso, R., et al.\
  2004, \apjl, 613, L153

\bibitem[Bundy \& Marcy(2000)]{2000PASP..112.1421B} Bundy, K.~A., \&
  Marcy, G.~W.\ 2000, \pasp, 112, 1421

\bibitem[Charbonneau et al.(2002)]{charbonneau02} Charbonneau, D.,
  Brown, T.~M., Noyes, R.~W., \& Gilliland, R.~L.\ 2002, \apj, 568,
  377

\bibitem[Eggenberger et al.(2004)]{2004A&A...417..353E} Eggenberger,
A., Udry, S., \& Mayor, M.\ 2004, \aap, 417, 353

\bibitem[Forbes(1911)]{1911MNRAS..71..578F} Forbes, G.\ 1911, \mnras,
  71, 578

\bibitem[Gaudi \& Winn(2007)]{gw07} Gaudi, B.~S., \& Winn, J.~N.\
2007, ApJ, in press [astro-ph/0608071]

\bibitem[Gimenez(2006)]{gimenez06} Gimenez, A.\ 2006, ApJ, 650, 408

\bibitem[McLaughlin(1924)]{mclaughlin24} McLaughlin, D.~B.\ 1924,
  \apj, 60, 22

\bibitem[Ohta et al.(2005)]{ohta05} Ohta, Y., Taruya, A., \& Suto, Y.\
  2005, \apj, 622, 1118

\bibitem[Ohta et al.(2006)]{ohta06} Ohta, Y., Taruya, A., \& Suto, Y.\
  2006, ApJ, submitted [astro-ph/0612224]

\bibitem[Queloz et al.(2000)]{queloz00} Queloz, D., Eggenberger, A.,
  Mayor, M., Perrier, C., Beuzit, J.~L., Naef, D., Sivan, J.~P., \&
  Udry, S.\ 2000, \aap, 359, L13

\bibitem[Rossiter(1924)]{rossiter24} Rossiter, R.~A.\ 1924, \apj, 60,
  15

\bibitem[Schlesinger(1911)]{1911MNRAS..71..719S} Schlesinger, F.\
  1911, \mnras, 71, 719

\bibitem[Snellen(2004)]{s04} Snellen, I.~A.~G.\ 2004, MNRAS, 353, L1

\bibitem[Ward \& Hahn (1994)]{wh1994} Ward, W. R., \& Hahn,
  J. M. 1994, Icarus, 110, 1

\bibitem[Ward \& Hahn (2003)]{wh2003} Ward, W. R., \& Hahn,
  J. M. 2003, AJ, 125, 3389

\bibitem[Winn, Holman, \& Roussanova(2006)]{whr06} Winn, J. N.,
  Holman, M. J., \& Roussanova, A.\ 2006, ApJ, in press
  [astro-ph/0611404]

\bibitem[Winn et al.(2005)]{wnhc05} Winn, J.~N. et al.\ 2005, ApJ,
  631, 1215

\bibitem[Winn et al.(2006)]{wjmb06} Winn, J.~N. et al.\ 2006, ApJ,
  653, L69

\bibitem[Winn et al.(2007)]{whhr07} Winn, J.~N. et al.\ 2007, AJ,
  submitted [astro-ph/0612224]

\bibitem[Wolf et al.(2007)]{wolf06} Wolf, A.S.\ et al.\ 2007, ApJ, in
  press

\bibitem[Wu \& Murray(2003)]{wu03} Wu, Y., \& Murray, N.\ 2003, \apj,
  589, 605

\end{thebibliography}
\end{document}